\documentclass[eqsecnum,floatfix,aps,prb,twocolumn,groupedaddress]{revtex4}

\usepackage{graphicx}

\begin{document}
\bibliographystyle{apsrev}

\title{Friedel Oscillations and Charge Density Waves in Chains and Ladders}

\author{Steven R. White}
\email{srwhite@.uci.edu}
\affiliation{Department of Physics and Astronomy, University of
California, Irvine, CA 92697}

\author{Ian~Affleck}
\email{affleck@physics.bu.edu}
\altaffiliation{On leave from Canadian Institute for 
Advanced Research and Department of Physics
 and Astronomy, University of British Columbia, Vancouver,
BC,  Canada, V6T 1Z1}
\affiliation{Physics Department, Boston University, 590 Commonwealth 
Ave., Boston, MA02215}

\author{Douglas J.~Scalapino}
\email{djs@vulcan.physics.ucsb.edu}

\affiliation{\sl Department of Physics,
         University of California,
         Santa Barbara, CA 93106-9530 USA}

\date{\today}
\begin{abstract}
The density matrix renormalization group method for ladders works much more
efficiently with open boundary conditions.  One consequence of these 
boundary conditions is
groundstate charge density
oscillations that often appear to be nearly constant in magnitude or to decay
only slightly away from the boundaries.
 We analyse these using bosonization
techniques, relating their detailed form to the correlation exponent and
distinguishing boundary induced generalized Friedel
oscillations from true charge density waves.  We also discuss a different
approach to extracting the correlation exponent from the finite size
spectrum which uses exclusively
open boundary conditions and can therefore take advantage of data for
much larger system sizes.  A general discussion of the Friedel
oscillation wave-vectors is given, and a convenient Fourier transform
technique is used to determine it.  DMRG results are analysed on
Hubbard and $t-J$ chains and 2 leg $t-J$ ladders.
We present evidence for the existence of a long-ranged
charge density wave state in the $t-J$ ladder at a filling of
$n=0.75$ and near $J/t \approx 0.25$.
\end{abstract}

\maketitle

\section{introduction}
Bosonization analyses\cite{Fabrizio,Schulz,Balents,Orignac}
 together with finite size numerical work using
the DMRG\cite{Noack,Siller} and other methods\cite{Hayward}
 have given a clear understanding of the
behavior of the
 2-leg Hubbard and $t-J$ ladder models.  The Hubbard Hamiltonian is
written:
\begin{eqnarray}
&H&= -t\sum_{i, \lambda, \alpha} \left(c^\dagger_{i+1\lambda\alpha}
c_{i\lambda\alpha} + h.c.\right)\\ 
&-& t\sum_{i,\alpha}
\left(c^\dagger_{i2\alpha}c_{i1\alpha}+h.c.\right) 
+ U\sum_{i,\lambda}
n_{i,\lambda ,\uparrow}n_{i,\lambda ,\downarrow}\nonumber
\label{HHubb}
\end{eqnarray}
Here $c_{i\lambda\alpha}$ destroys an electron on rung $i$ and leg
$\lambda=1,2$ with spin $\alpha=\uparrow,\downarrow$.
$n_{i,\lambda ,\sigma}$ is the electron number operator.
The $t-J$ Hamiltonian is:
\begin{eqnarray}
H= &-&t\sum_{i, \lambda, \alpha} \left(c^\dagger_{i+1\lambda\alpha}
c_{i\lambda\alpha} + h.c.\right) \nonumber\\
&-& t\sum_{i,\alpha}
\left(c^\dagger_{i2\alpha}c_{i1\alpha}+h.c.\right)\nonumber\\
&+&J\sum_{i,\lambda}\left(\vec S_{i\lambda}\cdot \vec S_{i+1\lambda} -
\frac{n_{i\lambda}n_{i+1\lambda}}{4}\right)\nonumber\\
&+& J\sum_i\left(\vec S_{i1}\cdot\vec S_{i2}-\frac{n_{i1}n_{i2}}{4}\right)
\label{HtJ}
\end{eqnarray}
where  $\vec
S_{i\lambda}=c^\dagger_{i\lambda}\ \vec\sigma/2\ c_{i\lambda}$ and
the Hilbert
space now excludes all states with doubly-occupied sites.
We will generally set $t=1$ in what follows.
Both models are expected to be in a ``C1S0'' phase, over a wide range of
parameters, in which
the low energy degrees of freedom consist of a single free massless
charge boson, whose excitations carry even multiples of the 
electron charge.  This is often characterized as a ``d-wave superconductor''
based on the nature of the power law decay of pair correlations which 
have a positive sign for singlet rung-rung or leg-leg correlations and 
a negative sign for rung-leg correlations. 
Both bosonization and DMRG results exist on multi-leg ladders with
 analytical and numerical uncertainty which increase with the number of
legs.

Most of the DMRG data is obtained with open boundary conditions along
the legs of the ladder.  Unlike with periodic boundary conditions,
this generally leads to charge density oscillations in the groundstate.
If these oscillations persist at the center of the chain for arbitrarily
long chain length then they correspond to a charge density wave (CDW).
More commonly, they decay away from the boundaries with a power law,
in the limit of an infinite chain.  In this case, we may think of them
as generalized Friedel oscillations where the chain ends themselves act as
impurities which induce gradually decaying density oscillations.  Unlike
in a Fermi liquid, these generalized Friedel oscillations in a Luttinger liquid
decay with an exponent which depends on the interaction strength,
and which is simply related to the density correlation 
exponent.\cite{Eggert,Egger}
Furthermore, the wave-vector of the oscillations itself can be 
changed by the interactions.  
These density oscillations, for the largest systems studied, up to
eight legs,
have been identified with ``stripes''.\cite{White2}  An understanding of the
occurance of stripes in these models, whether or not they require
long range Coulomb interactions to exist and their connection
with superconducitivity are important open problems.  Since
any ladder system is ultimately one-dimensional if the number of
legs is held fixed and the length taken to $\infty$, the
Friedel oscillations /CDW analysis may provide the appropriate
description of stripe behavior, at least in this limit and in
cases where the stripe wave-vector is parallel to the chains.

In Sec. II we briefly review the bosonization picture of
the 2-leg $t-J$ ladder.  We also present a technique for extracting
the correlation exponent purely from data with open boundary
conditions which has not, to our knowledge, been used previously
on ladders, although it has been used on chains.  This method is
used to obtain  values for the charge velocity and exponent.
In Sec. III we review the bosonization treatment of
commensurate charge density waves and the related behavior of
the correlation exponent, and analyse data on the 2 leg tJ model
at electron density $n=3/4$, showing that a CDW may occur for weak 
enough $J$. In Sec. IV we review
Friedel oscillations in Luttinger liquids and apply this analysis
to data on the single chain Hubbard and $t-J$ model and the 2 leg
$t-J$ model.

\section{The 2-leg Ladder}
A convenient starting point for bosonization is the weak coupling
Hubbard model version of the 2-leg ladder, Eq. (\ref{HHubb}) with 
$U/t$ small. 
  It is then far from
obvious that this analysis will apply to the infinite coupling
limit, corresponding to the $t-J$ model of Eq. (\ref{HtJ}) 
so that comparisons with
numerical results is important.  In the weak coupling limit we
may start by diagonalizing the non-interacting problem, giving
symmetric and anti-symmetric electron operators $\psi_{\lambda \alpha}$
where $\lambda =e$, $o$ labels even and odd channels.  We pass
to the continuum limit by introducing left and right moving fields:
\begin{equation}
\psi_{\lambda\alpha}(x)=e^{-ik_{F\lambda}x} \psi_{L\lambda\alpha}(x) +
e^{ik_{F\lambda}x} \psi_{R\lambda\alpha}(x)
\label{five}
\end{equation}
with $k_{Fe}$ and $k_{Fo}$ the fermi wave
vectors for the two bands.
In the usual way, we represent the left and right moving fermion fields
by left and
right moving boson fields:
\begin{equation}
\psi_{L/R\lambda \alpha}\propto e^{i\sqrt{4\pi}\phi_{L/R\lambda \alpha}}
\end{equation}
and then, introducing the dual canonical Bose fields,
\begin{equation}
\phi_{\lambda\alpha}= \phi_{R\lambda\alpha} + \phi_{L\lambda\alpha}
\qquad \theta_{\lambda\alpha} = \phi_{R\lambda\alpha}-\phi_{L\lambda\alpha}
\label{nine}
\end{equation}
It is then convenient to introduce spin and charge bosons, for each
channel:
\begin{eqnarray}
\phi_{\lambda\rho}&=& (\phi_{\lambda\uparrow} +
\phi_{\lambda\downarrow})/\sqrt{2}\nonumber\\
\phi_{\lambda\sigma} &=& (\phi_{\lambda\uparrow} - \phi_{\lambda\downarrow}
)/\sqrt{2}
\label{eleven}
\end{eqnarray}
and then, finally, switch to the two linear combinations of the
even and odd bosons:
\begin{equation}
\phi_{\pm\rho} = \left(\phi_{e\rho} \pm \phi_{o\rho}\right)/\sqrt{2}
\label{thirteen}
\end{equation}
with similar relations for $\phi_{\pm\sigma}$, $\theta_{\pm\rho}$, and
$\theta_{\pm\sigma}$. Actually, this last transformation is not
canonical when the even and odd bosons have different velocities.
However, we will follow the standard practice \cite{Fabrizio,Balents}
 of assuming that this
velocity difference is irrelevant.

A renormalization group analysis, based on the weak coupling Hubbard
model, suggests that the cosine interactions ``pin'' the bosons
$\theta_{\pm \sigma}$ and $\phi_{-\rho}$, introducing excitation
energy gaps for these bosons and leaving $\phi_{+\rho}$ as the only
massless boson which thus describes the low energy excitations. All
interactions involving $\phi_{+\rho}$, $\theta_{+\rho}$ are irrelelvant
in this phase.  The low energy effective Hamiltonian can be written:
\begin{equation}
H-\mu N={v_{+\rho}\over 2}\int dx \left[
K_{+\rho}\Pi_{+\rho}^2 +{1\over K_{+\rho}}\left( {d\theta_{+\rho}\over
dx}\right)^2\right],\label{Ham}
\end{equation}
where $\Pi_{+\rho}$ is the momentum density variable canonically
conjugate to $\theta_{+\rho}$,
$v_{+\rho}$ is the velocity of the corresponding gapless low energy
excitations and the parameter $K_{+\rho}$ controls the correlation
exponents.  (Our definition of $K_{+\rho}$ corresponds to that 
of Schulz\cite{Schulz} and Hayward and Poilblanc\cite{Hayward}
but is the inverse of the parameter with the same name in 
Balents and Fisher.\cite{Balents})
 The Hamiltonian may be equally well written in terms of
the other boson field $\phi_{+\rho}$ and its conjugate momentum using:
\begin{equation} \Pi_{\theta} = d\phi /dx,\ \
\Pi_{\phi}=-d\theta /dx.\label{Pi}\end{equation}
 The two parameters, $v_{+\rho}$ and $K_{+\rho}$ are
generally difficult to calculate analytically and are extracted from
numerical data.  The long-range behaviors of various correlation
functions are calculated straightforwardly by expressing the
corresponding fermionic operators in terms of the bosons
$\phi_{\pm \sigma}$, $\phi_{\pm \rho}$ and their duals.  Exponentials
of the pinned bosons can be replaced by their groundstate expectation
values but exponentials of the duals of pinned bosons lead to
exponentially decaying factors in correlation functions.  Exponentials
of the gapless boson, $\theta_{+\rho}$ and its dual give power law
decaying factors.  For instance, to calculate the uniform part of
the pair correlation function we bosonize the pair operator:
\begin{eqnarray}
\Delta_e &\equiv& \psi_{Le\uparrow}\psi_{Re\downarrow}\propto
e^{i\sqrt{4\pi}(\phi_{Le\uparrow}+\phi_{Re\downarrow})}\nonumber \\
& \propto&
e^{i\sqrt{\pi}(\phi_{+\rho}+\phi_{-\rho}-\theta_{+\sigma}
-\theta_{-\sigma})}.\label{pair}
\end{eqnarray}
We may replace the exponentials of $\phi_{-\rho}$ and $\theta_{\pm
\sigma}$ by a constant factor leaving simply the operator
$e^{i\sqrt{\pi}\phi_{+\rho}}$.  The correlation function for this
operator decays as $1/|x|^{1/(2K_{+\rho})}$.  The same result is
obtained for the correlation function of a pair of electrons in
the odd channel, $\Delta_0$.  On the other hand, the correlation
function $<\Delta^\dagger_e(x)\Delta_o(y)>$ is the same except for
a factor of $<e^{i\sqrt{4\pi}\phi_{-\rho}}>$.  This is expected to
be $<0$, based on the sign of the cosine interaction which pins
the $\phi_{-\rho}$ boson to $\sqrt{\pi /4}$, 
so this pair correlation function has
the opposite sign, corresponding to ``d-wave pairing''.  The
$2k_F$ part of the ( even, spin-up) density operator is:
\begin{equation}
e^{-2ik_{Fe}x}\psi^\dagger_{Re\uparrow}\psi_{Le\uparrow}
\propto e^{-2ik_{Fe}x}e^{-i\sqrt{\pi}(\theta_{+\rho}+\theta_{-\rho}
+\theta_{+\sigma}+\theta_{-\sigma})}.\end{equation}
The correlation function of this operator has exponential decay
due to the $e^{-i\sqrt{\pi}\theta_{-\rho}}$ factor.  Exponential
decay is also obtained for all other terms in the $2k_F$ part of
the density operator.  On the other hand, if we consider the
correlation function for the square of the density operator, we
get power-law decay for the $4k_F$ part.  This arises from terms of the form:
\begin{eqnarray}
&e^{-2i(k_{Fe}+k_{Fo})x} \psi^\dagger_{Re\uparrow}\psi_{Le\uparrow}
\psi^\dagger_{Ro\uparrow}\psi_{Lo\uparrow} \propto\nonumber\\
&\qquad e^{-2i(k_{Fe}+k_{Fo})x}e^{-2i\sqrt{\pi}(\theta_{+\rho}+
\theta_{+\sigma})} .
\label{density}
\end{eqnarray} 
We may replace
$\exp ( -2i\sqrt{\pi}\theta_{+\sigma})$ by its
expectation value
leaving only the gapless $\theta_{+\rho}$ field which gives power
law decay:
\begin{equation}
<n(x)^2n(0)^2>\propto {\cos [2(k_{Fe}+k_{Fo})x+\alpha ]\over
|x|^{2K_{+\rho}}}.\end{equation} In fact we expect that the
density operator itself will pick up a $4k_F$ term proportional to
this and also exhibit the same type of $4k_F$ power law decay.
This phase with one gapless charge mode and no gapless spin modes,
labeled ``C1S0'' is expected to be the generic phase in the 2 leg
$t-J$ or Hubbard ladder.  Whether this phase is better thought of as
almost a d-wave superconductor, or almost an incommensurate CDW
depends on the magnitude of $K_{+\rho}$.  The pairing correlation
function decays less rapidly than the density correlation function
in the case $K_{+\rho} >1/2$.  Schulz has shown\cite{Schulz2}
 that $K_{+\rho}\to 1$ 
as the density, $n$, approaches $1$.  At $n=1$, the ladder is a 
spin gapped Mott insulator.  

Thus it is of interest to calculate $K_{+\rho}$.  This can be done
from the finite size spectrum using either periodic or open
boundary conditions.  The latter are often used with DMRG so
we discuss both cases.  While, as we discuss below, 
 it is enough to measure the excitation energies
of only 3 states to extract $K_{+\rho}$ (and the velocity
$v_{+\rho}$) we give formulas for an infinite number of low lying
excitations.  This allows for a more extensive check of the
bosonization results with DMRG.  Due to the somewhat conjectural
nature of the extrapolation of weak coupling Hubbard results to
the $t-J$ model this would be a worthwhile check.  We discuss only the
groundstate and low-lying excitations, whose excitation energies
all scale to 0 as $1/L$, where $L$ is the length of the system.
These can be simply calculated from the free boson Hamiltonian of
Eq. (\ref{Ham}) when proper account is taken of the boundary
conditions on the boson field, $\theta_{+\rho}$.   We will
 generally set $v_{+\rho}=1$, restoring it by dimensional
analysis when needed.

We first consider the case of periodic boundary conditions
(b.c.'s).  Clearly  periodic b.c.'s on the fermions
translate into periodic b.c.'s on the boson fields but we must
take into account the fact that the boson fields are actually
phase fields (only their exponentials and derivatives occur as
physical local operators).  Since all physical operators in the
low energy effective theory involve integer powers of $\exp
[i\sqrt{\pi}\phi_{+\rho}]$ and $\exp
[i2\sqrt{\pi}\theta_{+\rho}]$, as in Eqs. (\ref{pair}) and
(\ref{density}), we see that:
\begin{eqnarray}
\phi_{+\rho} (L)&=&\phi(0)+2\sqrt{\pi}m\nonumber \\
\theta_{+\rho} (L)&=&\theta (0)+\sqrt{\pi}p,\label{bc}
\end{eqnarray}
where $m$ and $p$ are arbitrary integers.  To obtain the complete
low energy spectrum we simply write a mode expansion for the boson
fields, consistent with these b.c.'s and the canonical commutation
relations. This gives:
\begin{eqnarray}
&\phi_{+\rho}(x)& = \phi_0+{2\sqrt{\pi}mx\over L} \nonumber\\
&+& \sum_{k=1}^\infty
\sqrt{1\over K_{+\rho}4\pi k}\left(
a_{Rk}e^{i2\pi kx/L}+a_{Lk}e^{-i2\pi kx/L}+h.c.\right) \nonumber \\
&\theta_{+\rho}(x)&= \theta_0+ {\sqrt{\pi}px\over L} \\
&+& \sum_{k=1}^\infty
\sqrt{K_{+\rho}\over 4\pi k}\left( a_{Rk}e^{i2\pi
kx/L}-a_{Lk}e^{-i2\pi kx/L}+h.c.\right) \nonumber.\label{mode}
\end{eqnarray}
Here the quantum numbers $m$ and $p$ are integer valued.  Using
Eq. (\ref{Pi})  we see that the zero wave-vector mode
of the conjugate momentum, $\Pi_{\theta}$,
 has eigenvalues $2\sqrt{\pi}m$.  Thus
the wave-functions for the zero-momentum modes are $\exp [
i2\sqrt{\pi} m\theta_0]$.  These wave-functions are invariant
under $\theta_0 \to \theta_0 + \sqrt{\pi}$, as required by the angular
nature of $\theta$. (The same result holds for the dual zero wave
vector field, $\phi_0$ and its conjugate momentum.) The $a_{LK}$
and $a_{Rk}$ operators annihilate left and right moving harmonic
boson modes. We may give a physical interpretation to the quantum
number $m$ using the bosonization formula for the total charge
density:
\begin{eqnarray} J &\equiv& \sum_{\lambda ,\alpha}
(\psi^\dagger_{L,\lambda ,\alpha}\psi_{L,\lambda ,\alpha}
+\psi^\dagger_{R,\lambda ,\alpha}\psi_{R,\lambda ,\alpha})
\nonumber \\ &&\to (-1/\sqrt{\pi})\sum_{\lambda
,\alpha}d\theta_{\lambda ,\alpha}/dx\nonumber \\
&=&(-2/\sqrt{\pi})d\theta_{+\rho}/dx.\end{eqnarray} Thus the total
charge, relative to that of the groundstate, is:
\begin{equation} Q = (-2/\sqrt{\pi})\int_0^Ldx d\theta_{+\rho} /dx =-2\Delta
\theta_{+\rho} /\sqrt{\pi} = -2p.\label{Q} \end{equation} Only
 excitations of even charge occur in the C1S0 phase.  The other quantum
number, $m$, measures the ``chiral charge'' which has a less
obvious physical interpretation as the difference of charges of
left and right movers.

Inserting Eq. (\ref{mode}) into
the Hamiltonian of Eq. (\ref{Ham}) and using the identity, Eq.
(\ref{Pi})
we can immediately read off the
spectrum:
\begin{widetext}
\begin{eqnarray}
E-E_0=-2p\mu + {2\pi v_{+\rho}\over L}\left[K_{+\rho}m^2
+{p^2\over 4K_{+\rho}} 
+\sum_{k=1}^\infty k(n_{Lk}+n_{Rk})\right].
\label{Es}
\end{eqnarray} 
\end{widetext}
Here $n_{Lk}$ and
$n_{Rk}$ are the occupation numbers for the left and right moving
states of momentum $\pm 2\pi k/L$.  $E_0$ is the groundstate energy 
for a given density, $n$, and is non-universal.  This formula 
gives the excitation energy for low-lying excitations with all 
quantum numbers $\ll L$.  The parameters $\mu$ (chemical potential) 
$v_{+\rho}$ and $K_{+\rho}$ all depend on density.

The parameters $K_{+\rho}$ and $v_{+\rho}$ can be determined by
measuring the excitation energies of three states.  Generally one
calculates the excitation energies of the states with $-\Delta Q/2=p=\pm
1$ (and all other quantum numbers set to zero) to determine the
ratio $v_{+\rho}/K_{+\rho}$, using:
\begin{equation}
E(p=1)+E(p=-1)-2E_0 = {\pi v_{+\rho}/(K_{+\rho} L)}.
\label{detK}\end{equation} The compressibility for a 2-leg
ladder is generally defined as 
\begin{equation}{1\over n^2\kappa} \equiv {1\over 2L}{d^2E\over dn^2},
\end{equation}
 where $n$ is the density.  From Eqs. (\ref{Q}) and (\ref{Es}), this has
the value $\pi v_{+\rho}/(2K_{+\rho})$. The velocity may be measured
separately from the excitation energy of the lowest state of
momentum $2\pi /L$:
\begin{equation}
E(n_{R1}=1)-E_0 = {2\pi v_{+\rho}\over L}.\label{detv}\end{equation}
Thus measuring three excitation energies, for fixed, large $L$,
allows a determination of the correlation exponents.  Using a
large $L$ is important because corrections to Eq. (\ref{Es}) are
only down by additional powers of $1/L$.  These corrections become
especially large at commensurate filling near a transition to a
CDW.  The corresponding exponent will be given in Sec. III.  While
these three measurements are enough to determined the critical
exponents, measurements of energies of additional states and a
study of the $L$-dependence provide additional confirmation of the
predictions of the RG and bosonization.

As mentioned above, DMRG works much more efficiently with open
boundary conditions. We now discuss the finite size spectrum in
that case.  If we number the sites from 1 to L then we have a free
boundary condition at $j=1$ and $j=L$. A free boundary condition
on the fermion field at $j=1$ is equivalent to a vanishing
boundary condition at $j=0$. [This follows from adding an extra
``phantom site'' at $j=0$ with associated hopping and exchange
terms but then making these terms vanish by imposing
$\psi_{\lambda \alpha}(0)=0$.]  In terms of left and right movers
this free boundary condition becomes:
\begin{equation}
\psi_{L\lambda \alpha}(0)=-\psi_{R\lambda
\alpha}(0).\end{equation} Upon bosonizing we obtain:
\begin{equation}
e^{i\phi_{L\lambda \alpha}}=-e^{i\phi_{R\lambda
\alpha}},\end{equation} or
\begin{equation}
\theta_{\lambda
\alpha}(0)=\hbox{constant}.\label{bc1}\end{equation} (A
determination of this constant involves consideration of some
subtle commutators.  We will not bother to keep track of it in
what follows.)  Clearly this boundary condition remains the same
when expressed in the eventual basis $\rho /\sigma$, $+/-$.  The
only important boundary condition for the low energy excitations
is:
\begin{equation}
\theta_{+\rho}(0)=\hbox{constant}.\label{obc}\end{equation} The same b.c. is
obtained at $x=L+1\approx L$ except that the constant will
generally be different.  Taking into account the periodic nature
of $\theta$, Eq. (\ref{bc}), and the fact that
$\Pi_{\theta}\propto \partial \theta /\partial t\propto \partial \phi /\partial
x$ must also vanish at the boundaries,  we see that the 
mode expansions become:
\begin{eqnarray}
\phi_{+\rho} (x) &=& \phi_0 + \sum_{k=1}^\infty
\sqrt{1\over K_{+\rho}\pi
k}\cos \left({ \pi kx\over L}\right) \left(
a_{k}+a_k^\dagger \right) \nonumber \\
\theta_{+\rho}(x)&=& \theta_0+ {\sqrt{\pi}(p-\alpha /2\pi )x\over L}
\nonumber\\
&+& \sum_{k=1}^\infty \sqrt{K_{+\rho}\over \pi k}i\sin \left({ \pi
kx\over L}\right) \left( a_{k}-a_k^\dagger \right)
.\label{modeopen}
\end{eqnarray}
Here $\alpha$ is proportional to the difference of constants
appearing in the b.c.'s at $x=L$ and $x=0$.  Substituting into the
Hamiltonian, we now obtain the finite size spectrum:
\begin{eqnarray}
E-E_0=-2p\mu  +{\pi v_{+\rho}\over L}\left[
{(p-\alpha /2\pi )^2\over 2K_{+\rho}}+\sum_{k=1}^\infty
kn_{k}\right].\label{Esopen}\end{eqnarray} We see that Eq.
(\ref{detK}) for $v_{+\rho}/K_{+\rho}$ remains true with free b.c.'s
as we might expect due to the relation with the compressibility.
We may now determine the velocity independently (and hence
determine $K_{+\rho}$) by measuring the gap to the first excited
state with the same charge as the groundstate:
\begin{equation}
E(n_1=1)-E_0={\pi v_{+\rho}\over L}.\label{detvopen}\end{equation}
Note that this gap has half the value of the gap to the lowest
energy state of momentum $2\pi /L$ in the case of periodic b.c.'s,
Eq. (\ref{detv}). This is just a consequence of the familiar
result that the spacing of wave-vectors for open b.c.'s is $\pi
/L$, half the spacing for periodic b.c.'s.    

Previous measurements of the parameter $K_{+\rho}$, and its 
analogue in other chain and ladder systems, have
used periodic boundary conditions.  Exact diagonalization, 
or ``modified Lanczos'' methods work efficiently to 
determined the lowest energy state of given quantum numbers. 
Thus the energy differences of Eqs. (\ref{detK}) and (\ref{detv}) 
can be measured from finding the lowest energy states with 
various charges and with momenta $0$ and $2\pi /L$.  An 
alternative approach is to meaure the dependence of the 
groundstate energy (for a fixed charge) on an applied flux, 
i.e. a twist in the boundary conditions on the fermion fields:
\begin{equation}
\psi_{L/R\lambda \alpha}(L)=e^{i\Phi}\psi_{L/R\lambda \alpha}(0).
\end{equation}
  This corresponds 
to putting a twist into the boundary condition on $\phi_{+\rho}$ 
in Eq. (\ref{bc}), as can be seen from the fact that all 
the fermion fields ($L$ or $R$) contain a phase factor of 
$\exp [i\sqrt{\pi}\phi_{+\rho}/2]$. 
\begin{equation}
\phi_{+\rho}(L)=\phi_{+\rho}(0)+2\sqrt{\pi}(m+2\Phi ).\end{equation}
From Eq. (\ref{Es}) we see that the groundstate energy is 
increased by:
\begin{equation}
E_0\to E_0+8\pi v_{+\rho}K_{+\rho}\Phi^2/L.\end{equation}
Hence measuring the flux dependence of the groundstate 
energy determines $v_{+\rho}K_{+\rho}$, while 
measuring the compressibility determines $v_{+\rho}/K_{+\rho}$. 
Hayward and Poilblanc\cite{Hayward} measured $K_{+\rho}$ in 
this way for the 2-leg ladder using a maximum system size of 
$2\times 10$.  Another approach was taken by Siller et al.\cite{Siller}
 They modeled the 2-leg tJ model as a single chain model of 
bosonic hole pairs.  The parameters in the effective Hamiltonian 
for hole pairs were determined from DMRG calculations with 
open boundary conditions on system sizes up to $40 \times 2$ 
for systems containing 2 holes or 4 holes only.  The resulting bosonic 
model was then studied using exact diagonalization (Lanczos) 
methods and periodic boundary conditions.  Since the 
number of effective bosons (of density $x/2$ where $x\equiv 1-n$) 
is small near 1/2-filling, it was possible to diagonalize large 
systems (up to 220 sites in the case of only 2 bosons).  

\begin{figure}[b]
\includegraphics[width=8cm]{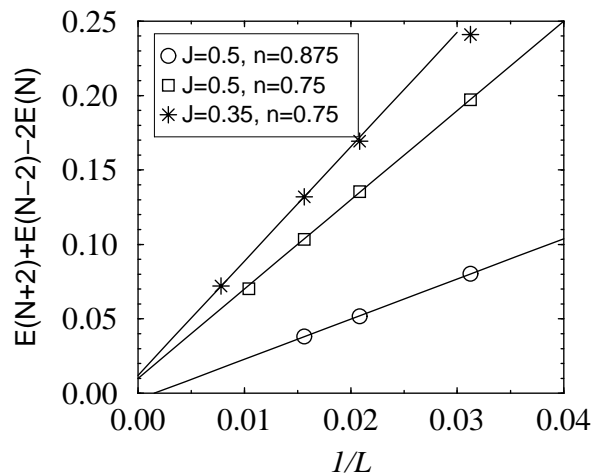}
\caption{These results were obtained by targetting 
three different ground states
with varying numbers of particles.
The straight lines are linear
fits to the data.
The slopes
of the fits (which asymptotically yield $\pi v /K$) are
2.70 for J=.5, n=0.875, 
6.75 for J=.5, n=0.75 and
7.68 for J=.35, n=0.75.}
\label{figdencgall}
\end{figure}

The approach advocated here, working exclusively with 
open boundary conditions, has been used previously for 
spin chains\cite{Eggert2} but not, as far we we know, 
for tJ ladders.  It allows us, using DMRG, 
 to study much larger system sizes, up 
to $192 \times 2$,  
than are accessible with Lanczos in the usual 
formulation.\cite{Hayward}  It furthermore avoids making 
any assumptions which are necessary in treating
the hole pairs as bosons, such as the particular form of the boson-boson
interaction and the absence of three-boson terms, which could induce some
density dependence in the interaction. 
However, the DMRG calculations with the accuracy and system length
required to determine $K_{+\rho}$ for these ladder systems were
surprisingly difficult. In order to minimize the effect of
correction terms to the asymptotic formulas, long systems were
needed. In these systems, one is probing competing pairing and
charge fluctuations at very low energies and large distances,
which is difficult in DMRG. Typically more than 2000 states were
kept in the calculations and more than a dozen sweeps were
performed. Some of the calculations were more accurate and
reliable than others, however.
In Fig. \ref{figdencgall} we 
show a plot of the left-hand side of Eq. (\ref{detK}) versus 
$1/L$ for three values of $(J,n)$, showing linear behavior and 
allowing us to extract $v_{+\rho}/K_{+\rho}$.   These
calculations require only the ground state energy, for which
an extrapolation in the energy versus the truncation error is
very reliable. Thus this determination of $v_{+\rho}/K_{+\rho}$
is the most reliable and accurate ingredient in the
determination of $K_{+\rho}$, and, in fact, the results shown
could be improved with modest additonal effort.

\begin{figure}[t]
\includegraphics[width=8cm]{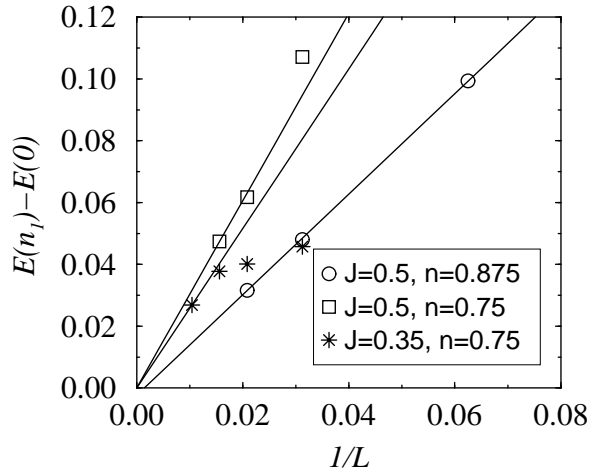}
\caption{Excitation energy for lowest state with $S^z=0$ and same 
electron density. 
The spin gap  is about 0.14
for J=0.5, n=0.875, about 0.10 for J=.5, n=0.75 and about
0.07 for J=.35, n=0.75. The breakdown of the asymptotic linear behavior is
 clearly visible in the third case, at energies of order the spin gap.
  The straight lines are linear fits to the data. 
The slopes
of the fits (which asymptotically yield $\pi v$) are 1.631 for n=0.875, J=0.5;
3.03 for n=0.75, J=0.5; and
1.6368 for n=0.75, J=0.35.}
\label{figcgall}
\end{figure}

In Fig. \ref{figcgall} we plot the left-hand side of Eq. (\ref{detvopen}) 
versus $1/L$, again obtaining linear behavior. These
calculations were difficult. It is necessary to target two
states simultaneously, and extrapolation is not very useful.
Note that, if $L$ 
is not large enough and the (infinite $L$) spin gap is not small 
enough, $\Delta_s<2\pi v_{+\rho}/L$, then the first excited state 
with the same quantum numbers might actually have $S=1$, $S^z=0$ 
and correspond to a spin excitation, rather than the desired 
neutral excitation of the $+\rho$ field.  (It might also be a neutral 
excitation of one of the other gapped boson fields.)  In order 
to check the former possibility, we have calculated the spin gap, 
i.e. the excitation energy for the lowest state with $S^z=1$.
We find that the very nonlinear behavior for $J=0.35$, $n=0.75$
shown in Fig. \ref{figcgall} is due to a small spin gap in
this system, of order 0.04. Thus it was necessary to study a
$96\times2$ system, for which we kept 4000 states, and performed
a dozen iterations, to clearly see the required gap. We believe
this data is reliable, but due to the large numerical work
required we have only studied three different values of $(J,n)$.
Combining the results of Figs. \ref{figdencgall} and \ref{figcgall}
allows a determination of 
$v_{+\rho}$ and hence $K_{+\rho}$.  The resulting values are 
given in Table \ref{table:v,K}.  

\begin{table}[t]
\begin{tabular}{r|r|r|r}
$(n,J)$ & $v$ & $K_{energy}$ & $K_{amp}$\\ 
\colrule 
$(0.875,0.5)$&$0.519$&0.604&0.633\\
$(0.75,0.5)$&$0.964$&0.449&0.359\\
$(0.75,0.35)$&$0.796$&0.33&0.284
\end{tabular} 
\label{table:v,K}
\caption{Results for the charge velocity $v$ and Luttinger liquid 
parameter $K$. $K_{energy}$ comes from using Eqs. (\ref{detK})
and (\ref{detvopen}), 
while $K_{amp}$ is determined using the decay of the Friedel
oscilations in the center of the system as a function of the
system length.}
\end{table}

\section{Charge density waves}
While a C1S0 phase is expected over most of the parameter range in
the 2-leg $t-J$  model, a completely gapped, charge density
wave (CDW)  phase
may occur at special commensurate filling factors, depending on
the value of $J/t$ (and other possible interaction 
parameters).\cite{Schulz3,Giamarchi}  In
this section we review the conditions on the critical exponent
parameter, $K_{+\rho}$ for a CDW to occur and discuss numerical
work on the $t-J$ model at electron densities $n=3/4$ and $n=1/2$,
corresponding to fillings of 3/8 and 1/4 respectively.

The low energy effective Hamiltonian only contains Fourier modes
of the electron fields within a small momentum range, $\pm \Lambda$
of $\pm k_F$. Umklapp type interaction terms, which do
not conserve separately the number of left and right movers, are
generally accompanied by rapidly oscillating phase factors.  They
therefore do not appear in the low energy effective Hamiltonian
since the electron fields vary slowly and hence these rapidly
oscillating factors cause these interactions to
average to zero.  However, at special filling factors, corresponding
to special values of the Fermi wave-vector, the oscillating factors
become constants and these operators then appear in the effective
Hamiltonian.  Whether or not they produce a gap and a CDW depends
on whether or not they are relevant operators in the RG sense.
In a 1-dimensional relativistic quantum field theory an operator
is relevant if it has a scaling dimension, $x<2$.  This scaling
dimension determines the exponent, $\eta$ with which the correlation
function of this operator decays, with:
\begin{equation}
\eta = 2x.\end{equation}
Using the free boson Hamiltonian of Eq. (\ref{Ham}), the scaling
dimension of any operator is easily obtained.
For example, the Hubbard or $tJ$ interaction leads to an interaction
term of the form given in Eq. (\ref{density}).  We see that
this oscillates at wave-vector $2(k_{Fe}+k_{Fo})$.  Only when this
wave-vector is a mutiple of $2\pi$ will this operator appear in
the low energy effective Hamiltonian.  Furthermore, any operators
whose correlation functions decay exponentially are irrelevant.

The number of electrons, $N_{\lambda}$ in each band is:
\begin{equation}
N_\lambda /L = 2k_{F\lambda}/\pi ,\end{equation}
(The factor of 2 arises from spin.) Thus:
\begin{equation}
2(k_{Fe}+k_{Fo}) = 2\pi n,\label{k-n}\end{equation}
where:
\begin{equation}
n \equiv (N_e +N_o)/2L,\end{equation}
is the electron density.  
While the value of $k_{Fe}$ and $k_{Fo}$ may be renormalized
by interactions (and may, in fact, not really be well-defined
in the interacting model) their sum is ``protected'' by the
one-dimensional version of Luttinger's theorem\cite{Yamanaka} so that
Eq. (\ref{k-n}) is expected to be exact.

Thus we see that the Umklapp operator of Eq. (\ref{density}) can
only occur in $H_{eff}$ for $n=1$ (half-filling).  This operator
has scaling dimension:
\begin{equation} x= K_{+\rho}.\end{equation}
At 1/4 filling, corresponding to $n=1/2$, the product of this operator times
itself with spin up replaced by spin down, can occur, giving rise to 
an operator
containing 8 electron operators.  Under bosonization this
is proportional to $\exp [-4i\sqrt{\pi}\theta_{+\rho}]$ and has
dimension $x=4K_{+\rho}$.  At a filling of 3/8 (or 1/8) corresponding 
to n=3/4 (or 1/4)  the fourth
power of the basic Umklapp operator in Eq. (\ref{density})
containing 16 electron operators can occur.  It has dimension
$x=16K_{+\rho}$.

The condition on the RG scaling dimension, $x$, for an operator
to induce a CDW is somewhat subtle, and depends on whether the
density or $J/t$ is varied.  If we vary $J/t$, holding the density
fixed at the commensurate value, then the system will remain
in the gapless C1S0 phase as long as $x>2$, so that the
operator is irrelevant.  A transition to a gapped CDW phase occurs
at the value of $J/t$ where $x$ becomes $<2$.  In the CDW phase
the value of $K_{+\rho}$ becomes undefined since all correlation
functions decay exponentially (or go to constants).  We may
also consider what happens as we vary the density in the vicinity
of the commensurate value.  If the system is not in the CDW phase
at a commensurate filling, then we expect $x$ to vary smoothly and,
of course, to  satisfy $x>2$ at the commensurate point.  On the other
hand, if the system {\it is} in the CDW phase at a commensurate
filling, then we expect $x\to 1$ as the commensurate filling is
approached.  This follows from a general and clever argument by
Schulz\cite{Schulz3} based on ``refermionizing'' the single boson Hamiltonian.
The relevant operator always behaves like a fermion mass term,
of scaling dimension $x=1$ very close to the commensurate filling,
whenever a CDW occurs at that point.  It follows that, $x$ must
vary rapidly in the vicinity of the critical value of $J/t$ where
the CDW transition occurs  near a commensurate filling.  On the
CDW side of the transition $x$ will be close to 2 away from the
commensurate filling but then drop abruptly to $1$ as that filling
is approached.  We emphasize that Schulz's argument, which was
formulated in terms of the single boson sine-Gordon theory,
is very general and doesn't depend on the underlying microscopic
model.  Similar behavior can occur for a $t-J$ or Hubbard ladder with
any number of legs.  Such rapid variation of $K$ in the vicinity
of commensurate filling for a system with a CDW was observed
by Schulz for the single leg Hubbard model near half-filling,
by obtaining $K$ from the Bethe ansatz and similar behavior 
was also observed near
1/2-filling ($n=1$) for the 2-leg $t-J$  model by Siller et al.

Now consider a density $n=1/2$ corresponding to 1/4 filling.  If we
sit at this density and vary $J/t$ then a CDW transition occurs
when $K_{+\rho}=1/2$, with the gapless phase having $K_{+\rho}>1/2$.
On the other hand, if $J/t$ is such that the system is in a CDW
phase at $n=1/2$, then $K_{+\rho} \to 1/4$ as $n\to 1/2$.

\begin{figure}
\includegraphics[width=8cm]{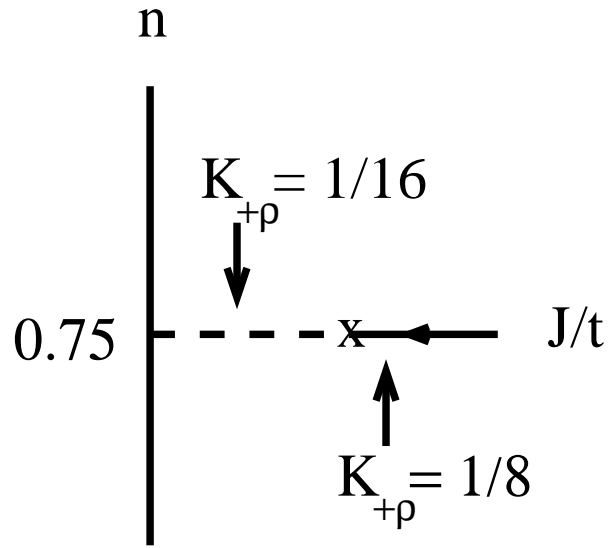}
\caption{The qualitative behavior of $K_{+\rho}$ as a function 
of $J/t$ and $n$.  A CDW occurs along the dashed line at $n=.75$ so
$K_{+\rho}$ is not well-defined there.  As this line is approached 
by varying $n$, $K_{+\rho}\to 1/16$.  On the other hand, at $n=.75$ 
and larger $J/t$ there is no CDW and $K_{+\rho}$ {\it is} well 
defined, with a value approaching $1/8$ at the CDW critical point.}
\label{fig:Krho}
\end{figure}

Similarly, a CDW at 3/8 filling, $n=3/4$, is signalled by
$K_{+\rho}\to 1/8$, as we vary $J/t$ at fixed $n$ or
$K_{+\rho}\to 1/16$ as we vary the density at fixed $J/t$ in the CDW
phase. See Fig. \ref{fig:Krho}. 

The connection between the relevance of the multiple Umklapp interaction
and the presence of a CDW is easily established.  When the multiple
Umklapp term (for $n=1/2$ or $3/4$) is relevant it pins the
$\theta_{+\rho}$ boson.  From Eqs. (\ref{density}) and (\ref{k-n})
  we see that the $4k_F$
term in the density operator then has a non-zero groundstate
expectation value, so that
\begin{equation}
<n_j> \propto \cos (2\pi nj+ \alpha )+\hbox{constant},
\label{densosc} \end{equation}
where $\alpha$ is a constant.  Thus the density oscillations
should have wavelength 2 for n=1/2 and wavelength 4 for n=3/4.  In
general, it is sometimes convenient to introduce
\begin{equation}
\delta \equiv 1-n,
\end{equation}
 the density of holes measured from half-filling, sometimes called
the density of ``holons'' and proportional to the doping parameter
in the cuprates.  We see that $n$ can be replaced by $\delta$ in
Eq. (\ref{densosc}) so that the wavelength of the CDW oscillations
is $1/\delta$.  For a 2-leg ladder, this is the average horizontal
separation of {\it hole pairs}.  (``Horizontal'' refers to the direction
along the legs of the ladder.) 

The CDW corresponds to a broken
translational symmetry, so that there are two different groundstates
for n=1/2 and four different groundstates for n=3/4 (differing
only by translation by 1 site).
For a finite ladder with periodic b.c.'s
we expect quantum tunnelling between these groundstates to occur
so that no density oscillations exist in the finite system groundstate.
On the other hand, with open b.c.'s one or the other of the groundstates
gets picked out by the boundary conditions so that the CDW is directly
observable.

We now return to the point raised in Sec. II, the higher order
corrections in $1/L$ to the finite size energy gaps given in Eqs.
(\ref{Es}) and (\ref{Esopen}).  The leading correction is
determined by the leading irrelevant operator.  At a commensurate
filling, when the system is close to having a CDW, the multiple
Umklapp operator discussed above has a dimension $x$ only slightly
greater than 2, corresponding to being barely irrelevant.  In this
case the higher order terms in the energy formula are of
$O(1/L^{x-1})$.  This will make it difficult to determined
$K_{+\rho}$ reliably from finite size gaps in the vicinity of a
CDW, complicating the determination of the CDW phase boundary.

\begin{figure}
\includegraphics[width=8cm]{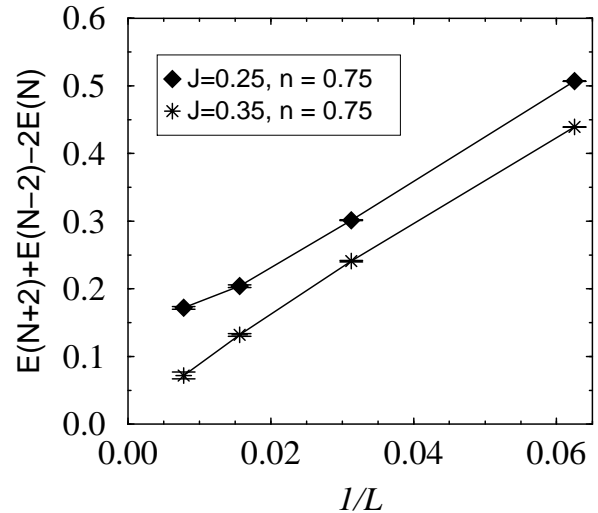}
\caption{Scaling of the charge gap with $1/L$, showing the occurance 
of a CDW at $n=.75$ for $J=.25$ but not $J=.35$.}
\label{figdencg.35.25}
\end{figure}

Previously reported work\cite{Siller} based on the bosonic 
hole pairs approximation together with DMRG and Lanczos, 
on the 2-leg $tJ$ model, with
$J/t=.35$,
found a smooth behavior of $K_{+\rho}$ in the vicinity of $n=3/4$
with a value at a commensurate density $n=3/4$ of approximately $.232$. 
Using the direct DRMG approach, we have found 
$K_{+\rho}\approx .33$.   Both results are
far above the critical values ($1/8$ and $1/16$),
indicating that a CDW does {\it not}
occur at $n=3/4$ for $J/t=.35$.  The charge gap for $n=.75$ with 
$J=.35$ and $J=.25$ is plotted versus $1/L$ in Fig. \ref{figdencg.35.25}.
There is, indeed, no evidence for a charge gap at the larger value of $J$, 
consistent with the absence of a CDW. 
On the other hand, there {\it is} evidence for a charge gap at 
$J=0.25$. Note that these calculations involve only ground state
energies and thus are very reliable. Although the results are
extrapolated to zero truncation error, the extrapolation is
quite small in magnitude, of order the symbol size in the
figure, and the estimated error in the extrapolation is very
small, as shown by the error bars.  These results provide some evidence
that there is a criticial value of $J$ between 0.25 and 0.35 such 
that a charge gap and CDW occur for smaller $J$.  

As reviewed in the next section, if there is no CDW, and the 
system in is the  C1S0 
phase, we expect the density oscillations at the center of the 
chain to decay with chain length as $L^{-K_{+\rho}}$ with 
$K_{+\rho}>1/8$.  Comparing the oscillation amplitude for 
$n=.75$, $J=.25$ and 
the two longest lengths studied, L=96 and L=192, we find 
a ratio of oscillation amplitudes of .915 which would give 
an exponent of $K_{+\rho}=.128$.  This is slightly 
larger than the critical value of 1/8.  On the other hand,
it is so close to the critical value that we might wonder 
if the density oscillation amplitude would eventually approach 
a constant with still longer chain lengths corresponding to 
a true CDW.  

\begin{figure}[b]
\includegraphics[width=8cm]{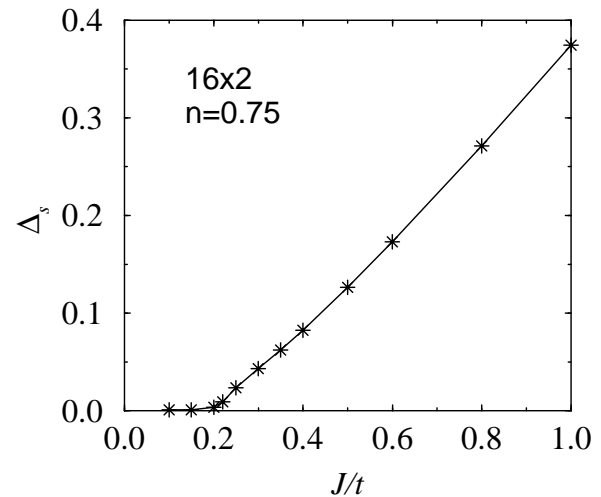}
\caption{Spin gap on a $16\times2$ system as a function of
$J/t$.}
\label{spingap16x2}
\end{figure}

Another important question is the behavior of the spin 
gap in the putative CDW phase.  Since the Umklapp 
term which drives the CDW does not contain the 
spin bosons, it is natural to assume that the 
gapping of the charge boson, corresponding to 
the CDW, occurs without any major effects on 
the other bosons which could thus remain gapped.  
However, it is possible that one (or more) of 
the other bosons, such as a spin boson, becomes 
gapless at the same transition.  
Indeed, the occurence of the CDW may be related to the close
proximity of a phase with no spin gap. 

\begin{figure}
\includegraphics[width=8cm]{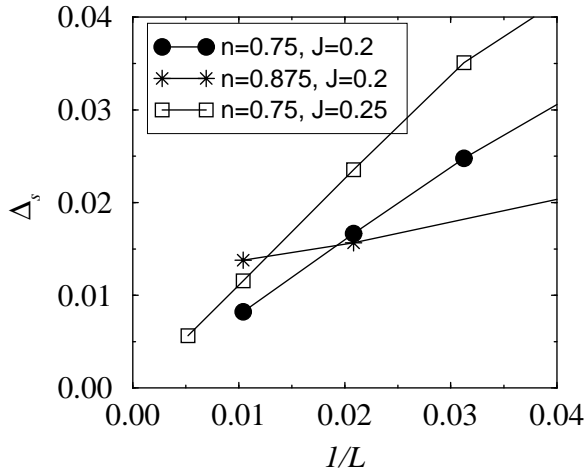}
\caption{Spin gap as a function of $1/L$ for several systems.
The values of the exchange on the first and last rungs have
been altered to avoid edge excitation.}
\label{spingapvsL}
\end{figure}

In Fig. \ref{spingap16x2},
we show the spin gap on a $16\times2$ system for a wide range
of $J$. The data suggests that in the thermodynamic limit,
the spin gap vanishes near $J=0.25$. To verify this result,
one must perform a finite size study. However, the spin 
density pattern on the $16\times2$ system for $J=0.2$ has the
largest values of $\langle S^z_i \rangle$ concentrated
near the ends of the system, indicating that an edge excitation
has the lowest energy. In order to ensure that we have
a bulk spin excitation in the calculation, we next studied
systems where we increased
the rung exchange interaction to $J+0.3$ on the first and
last rungs of a ladder. This drives up the energy of the
edge excitation, and one finds that the resulting spin
pattern is concentrated in the central region of the ladder.
In Fig. \ref{spingapvsL}, we show the spin gap as a function 
of $1/L$ for several systems. The results for $J=0.2$,
$n=0.75$ are consistent with a gapless state, although
one can never rule out a very small gap. Similarly, 
for $J=0.25$,
$n=0.75$ no evidence for a spin gap on the largest
system, a $192\times2$ ladder. For $J=0.2$, $n=0.875$ we see
a clear spin gap of order 0.013.

The proximity to a phase with vanishing spin gap may 
be related to the occurence of the CDW.  As the 
spin gap gets smaller, and the corresponding 
length scale larger, we may expect the size 
of the 2-hole pairs to get larger.  This is 
naturally associated with a growing scattering length 
for the effective boson system, which enhances 
the formation of a CDW.  

\begin{figure}
\includegraphics[width=4cm]{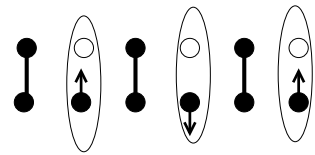}
\caption{A charge arrangement with period 2 at n=3/4, corresponding 
to a gapless CDW.  The black circles represent electrons and 
the white circles holes.  The ovals indicate that the location of the hole 
is symmetrized between the two sites on a rung.  The solid lines 
indicate the formation of dimer singlets. The arrows indicate the 
spins of the unpaired electrons, but we {\it donnot} expect them 
to be N\'eel ordered in this one-dimensional system.}
\label{fig:per2}
\end{figure}

\begin{figure}
\includegraphics[width=4cm]{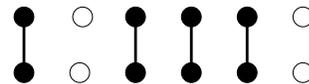}
\caption{A charge arrangement with period 4 at n=3/4, corresponding 
to a gapped CDW. (The actual density oscillations seem to correspond 
to pairs of holes shared evenly by every second pair of rungs.)}
\label{fig:per4}
\end{figure}

A CDW with vanishing spin gap at $n=3/4$, can be easily  understood 
heuristically. One can imagine one hole localized on every 
second rung, with the electrons on doubly occupied 
rungs forming spin singlets, as illustrated in Fig. \ref{fig:per2},
  giving an effective $S=1/2$ Heisenberg chain 
with lattice spacing 2 which has vanishing spin gap.  
On the other hand, in the spin-gapped phase we can think of 
pairs of holes localized on every fourth rung, as in fig. \ref{fig:per4}. 
 Thus we expect 
the groundstate density oscillations to have wavelength 2 in 
the CDW with vanishing spin gap but wavelength 4 in the 
gapped CDW. 

The connection between wavelength doubling and the spin gap is 
related to the Lieb-Schultz-Mattis theorem.\cite{Yamanaka} 
If the wavelength is not doubled to 4, then gapless excitations 
are expected.  More precisely, we can prove the existence of 
gapless excitations at n=3/4 if the groundstate is invariant 
under translations by 2 and under site parity: $j\to -j$ and is 
not ferromagnetic.   
(Note that {\it link} parity, $j\to -j+1$ {\it is} spontaneously 
broken by the wavelength 2 groundstate illustrated in Fig. \ref{fig:per2}
but site parity is not.)  This is proven by considering a long 
ladder of even length $L$  with {\it periodic} b.c.'s.  Let 
$|\psi_0>$ be a groundstate.  We then consider the variational 
state obtained by acting on the $|\psi_0>$ with the unitary 
operator:
\begin{equation}
U=\exp [i(2\pi /L)\sum_jjn_{j\uparrow}],\end{equation}
where $n_{j\uparrow}$ is the number operator for spin up electrons 
on rung $j$ (summing over both sites on the rung).  It is straightforward 
to calculate $<\psi_0|U^\dagger (H-E_0)|\psi_0>$ and show that 
it is $O(1/L)$ provided that the following term vanishes:
\begin{equation}
\sum_j[<\psi_0|c^\dagger_{j\lambda \uparrow}c_{j+1\lambda \uparrow}|\psi_0>
-<\psi_0|c^\dagger_{j+1\lambda \uparrow}c_{j\lambda \uparrow}|\psi_0>].
\end{equation}
We can show that this vanishes, for example, by using the 
assume site parity symmetry to show that:
\begin{equation}
<\psi_0|c^\dagger_{j\lambda \uparrow}c_{j+1\lambda \uparrow}|\psi_0>
=<\psi_0|c^\dagger_{j\lambda \uparrow}c_{j-1\lambda \uparrow}|\psi_0>.
\end{equation}
 To complete the proof of a low energy state
we must prove that $U|\psi_0>$ does not become $|\psi_0>$ in 
the limit $L\to \infty>$.  We prove this by considering the 
behavior of $U$ under translations by 2 sites, the assumed 
symmetry of the groundstate.  We fine that under this translation:
\begin{equation}
U\to Ue^{i8\pi n_{\uparrow}},\label{Utr}\end{equation}
where $n_{\uparrow}$ is the density of spin up electrons, 
$n_{\uparrow}=N_{\uparrow}/2L$, where $N_{\uparrow}$ is the total 
number of spin up electrons and $2L$ is the number sites on a 2-leg 
ladder of length $L$.  If we further assume that $n_{\uparrow} 
=n_{\downarrow}$, i.e. that the groundstate is not ferromagnetic, 
then we may replace the exponential factor in Eq. (\ref{Utr}) by 
$e^{i4\pi n}$.  For a density, $n=3/4$, this factor is -1.  This 
proves orthogonality of $U|\psi_0>$ with $|\psi_0>$ and hence 
the existence of a low energy excitation.  On the other hand, 
if the groundstate is only invariant under translations by 4 sites 
then, repeating the argument with a translation by 4 sites changes 
the exponential factor to $e^{i8\pi n}$ which is 1 for n=3/4.  
Thus we cannot prove that $U|\psi_0>$ is orthogonal to the 
groundstate in this case; it may approach it in the limit 
$L\to \infty$, so the proof collapses.  Unfortunately, this does 
not prove that there is a gap when the groundstate has wavelength 4, 
only that the gap neccessarily vanishes when it has wavelength 2.
 However, it often appears to be the case that when the LSM 
theorem fails, a gap appears.  for instance, for S=1/2 spin chains, 
a gap is expected whenever the groundstate has wavelenth 2, as in a 
dimerized state. Thus, the nature of the density oscillations 
provides additional evidence for which phase the system is in.  

To see what Fourier components are present in the density 
oscillations, we Fourier transform the density as a function
of rung position and plot the power spectrum. 
In order to avoid spurious edge effects, a windowing
function is used. If the original density is $d(x)$, we 
Fourier transform $D(x) = W(x) d(x)$, where $W(x)$ is a
smooth windowing function. $W(x)$ is chosen to vanish at $x=1$ and $x=L$,
and to be unity for the middle third of the $x$ values, with
a smooth continuation in between; the particular choice we use
is given in Ref. (\onlinecite{smooth}). Before Fourier transforming, we 
calculate the sum $I$ of $D(x)$ from $1$ to $L$,
and the sum $J$ of $W(x)$, and
then subtract from $D(x)$ the function $\frac{I}{J} W(x)$. This removes
a large peak centered at $k=0$ which is uninteresting. The
lattice spacing is set to 1, so that the exponential in the FT
is $\exp (-i k j)$, where j runs over integer lattice sites.
With this approach there is no restriction on the allowed values
of $k$; the finite value of $L$ instead leads to finite widths
for the peaks.  

\begin{figure}
\includegraphics[width=8cm]{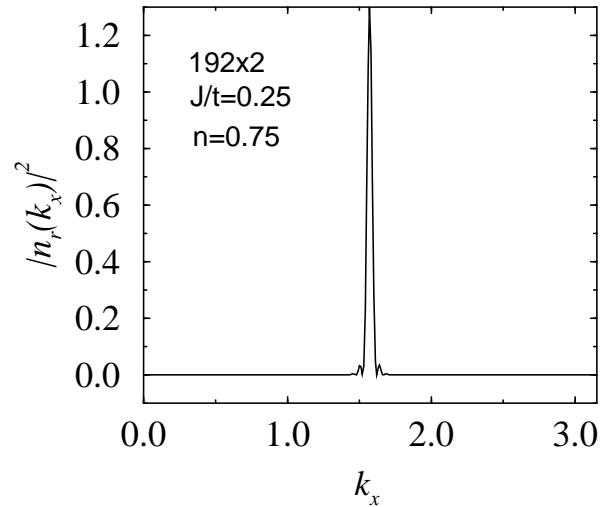}
\caption{Smoothed Fourier transform of the density oscillations 
showing only a large peak at wave-vector $\pi /2$ corresponding 
to a wavelength of 4.}
\label{192J.25}
\end{figure}

The density 
oscillations for $J=.25 $ and also $J=.2$, 
with $n=.75$ have a very strong component at
 a wavelength of 4, corresponding to a spin-gap, 
as shown in figure \ref{192J.25}.  

\section{A possible phase diagram}
\begin{figure}
\includegraphics[width=8cm] {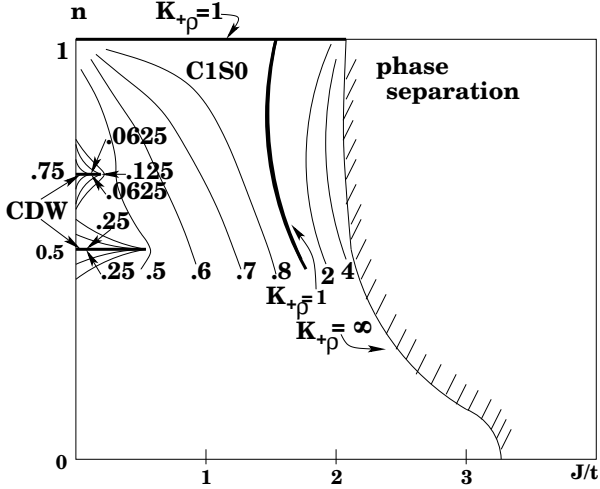}
\caption{Schematic phase diagram for the $t-J$ ladder 
as a function of $J/t$ and the electron density $n$ in the region 
$n$ greater than about .5.}
\label{fig:phase}
\end{figure}

In this section we briefly discuss a possible phase diagram 
and the qualitative behavior of $K_{+\rho}$ as a function 
of $J$ and $n$ 
for the 2-leg $t-J$ model. 
We are interested in the region $n\geq 0.5$ where both the even and odd parity
 bands have carriers in the weak coupling Hubbard limit. 
 At still lower doping the system is expected to enter
 a C1S1 phase. In the region $n \geq .5$ our phase diagram 
  is similar to that 
proposed by Hayward and Poilblanc,\cite{Hayward} and is based on their 
exact diagonalization results, earlier DMRG results, 
our new DMRG results and 
some general results which follow from bosonization and 
RG.  This is sketched in Fig. \ref{fig:phase}.
Here the two ordered CDW phases at n=0.75 and n=0.5 are indicated by solid
cuts for small values of J/t and a phase separation region occurs for
large J/t values. We conjecture that the remaining region is in a   
C1S0 (d-wave-4k$_F$ CDW) phase. As discussed in Sec. II, this phase has the
$\phi_{-\rho}$ and $\theta_{\pm \sigma}$ bosons pinned  and is characterized
by power law d-wave pairing and 4k$_F$ CDW correlations. When 
$K_{+\rho}>0.5$, the
pairing correlations are dominant. Supporting this conjecture, DMRG
calculations for J/t and n values near the phase separation boundary show
clear power law d-wave like pairing correlations.

The phase diagram shown in Fig. \ref{fig:phase} for the $t-J$ ladder differs in
several ways from that expected for the 2-leg repulsive U Hubbard model.
Firstly, it is generally believed that 
the Hubbard ladder doesn't exhibit phase separation and the ordered
CDW phases are absent. Secondly, weak coupling renormalization-group
studies, bosonization [4] and DMRG calculations suggest that 
$K_{+\rho}\leq 1$ for
the repulsive U Hubbard model.  For an extended Hubbard model with
additional interactions one can have $K_{+\rho} >1$;
however Orignac and Giamarchi\cite{Orignac} 
argue that when $K_{+\rho}>1$, the d-wave-4k$_F$ CDW sector gets replaced by a 
C1S0 s-wave pairing or an orbital antiferromagnetic phase. This raises
questions regarding the phase diagram show in Fig. \ref{fig:phase}. 
We believe that while
alternate C1S0 phases are certainly logical possibilities, they do not
occur in the $t-J$ ladder. The argument that they occur whenever $K_{+\rho}>1$
appears to depend upon a weak coupling analysis which may not be
applicable to the $t-J$ ladder. In this analysis, the phase boundaries
between the various C1S0 phases is determined by the equality of 2
small marginal coupling constants in the 
Hamiltonian.  These coupling constants also determine 
$K_{+\rho}$, in the weak coupling limit, in such a way 
that when they are equal, $K_{+\rho}=1$.  [See 
Eq. (35) of Ref. (\onlinecite{Orignac}).] For stronger 
coupling, and, in particular, for the $t-J$ model, 
the possible phase boundary to a phase with orbital 
antiferromagnetism need not be related in any general 
way to the value of $K_{+\rho}$ and the DMRG results 
suggest that orbital antiferromagnetism does not 
occur in the ordinary $t-J$ model.\cite{OAF} 

The phase separation boundary was obtained in a previous 
DRMG study.\cite{separation}
  As this boundary is approached the compressibility 
diverges and $K_{+\rho}\to \infty$.  

We have taken into account the behavior of $K_{+\rho}$ 
near commensurate fillings in drawing Fig. \ref{fig:phase}.  
In particular, at $n\to 1$, the $t-J$ model reduces 
to the 2-leg Heisenberg spin ladder and has an infinite 
gap for all charge excitations and presumably a gap of 
$O(J)$ for spin excitations.  The general arguments of 
Schulz\cite{Schulz2,Schulz} then imply that $K_{+\rho}\to 1$ along 
the line $n=1$.  We have also assumed that a CDW occurs 
at n=.75 and n=.5 for small enough $J$.  As discussed in 
the previous section, we have found some evidence for 
such a phase at $J=.25$ and n=.75.  The corresponding 
behavior of $K_{+\rho}$, reviewed in the previous section 
is incorporated in Fig. \ref{fig:phase}.  Note that, 
at small $J$, $K_{+\rho}$ apparently varies rapidly 
between .0625 for n=.75 and 1 for n=1.  As we also 
discussed in the previous section, it is  possible 
that the spin gap vanishes at commensurate fillings for 
some range of $J$.  We do not attempt to indicate this 
possibility in Fig. \ref{fig:phase}.  

For values of $J/t$ around .35, which has been argued 
to be a reasonable value for modeling the cuprates, 
$K_{+\rho}<.5$ near n=.75 but increases towards 1 as 
n=1 is approached.  Thus there is only a narrow window of 
doping near 1/2-filling where d-wave pairing correlations dominate.

\section{Friedel Oscillations}
At incommensurate filling, or in general when no CDW occurs, there
will still be density oscillations produced by the boundaries
of an open ladder which only decay slowly (with a power law) into
the ladder. Detailed predictions can be made about these using
bosonization generalizing the approach of Ref. (\onlinecite{Eggert,Egger}) 
for the spin chain and Hubbard chain 
 to the case of the ladder.
 These oscillations provide a further check on the theory.  Thus
consider again the $4k_F$ term in the density operator, of Eq.
(\ref{density}):
\begin{equation}
n_j= Ae^{-2i\pi nj}e^{-2i\sqrt{\pi}\theta_{+\rho}}+ h.c.\label{density2}
\end{equation}
where $A$ is a constant into which we have adsorbed
$<e^{-2i\sqrt{\pi}\theta_{+\sigma}}>$. As we discussed in Sec. II, 
this  $4k_F$ term in the continuum representation of $n_j$, 
determines the leading behavior of the density-density correlation 
function at long distances since any $2k_F$ terms decay exponentially.  
For an infinite system, the correlation
function of the exponential operator appearing in Eq. (\ref{density2}),
behaves as:
\begin{equation}
<e^{2i\sqrt{\pi}\theta_{+\rho}}(r)e^{-2i\sqrt{\pi}\theta_{+\rho}}(0)>
=c| r|^{-2K_{+\rho}},
\end{equation}
where $c$ is a constant, with dimensions of
(length)$^{2K_{+\rho}}$. Decomposing $\theta = \phi_R-\phi_L$, we
may write this correlation function as a square of two identical
factors, the correlation functions of $\exp
[2i\sqrt{\pi}\phi_{R/L}]$. It thus follows that the oscillating
term in the density correlation function is:
\begin{equation}
<n(r)n(0)> \to \cos (2\pi nr+\alpha ) 2c|A|^2 |r|^{-2K_{+\rho}}.
\label{dens-corr}\end{equation} Now consider the semi-infinite
system ($r>0$) with one free boundary condition. As remarked in
Sec. II, this corresponds to a b.c. on the boson field given in
Eq. (\ref{obc}), or equivalently:
\begin{equation}
\phi_L(t,0)=\phi_R(t,0)+\hbox{constant}.\end{equation} Now using
the fact that $\phi_R$ is a function of $vt-r$ only and $\phi_L$ a
function of $vt+r$ only, we see that this equation implies that we
may regard $\phi_R$ as the analytic continuation of $\phi_L$ to
the negative axis:
\begin{equation} \phi_R(r)=\phi_L(-r) +
\hbox{constant}.\end{equation} Thus the expectation value of
$n(r)$ reduces essentially to the square root of the correlation
function $<n(r)n(0)>$:
\begin{equation}
<n_j>\to 2\sqrt{c}|A|\cos (2\pi nj + \beta
)(2j)^{-K_{+\rho}}.\label{frie-inf}\end{equation} 
The wave-vector of the Friedel oscillations 
is the same as the wave-vector governing the long-distance 
oscillations in the correlation function.  No Friedel 
oscillations occur at wave-vector $2k_{Fi}$ as they would for 
the non-interacting system or more generally in phases where, 
for example, all 4 bosons are gapless. The exponent
governing the decay of the Friedel oscillations is $K_{+\rho}$,
1/2 the exponent governing the decay of density correlations.
Furthermore, the amplitude is simply the square root of the
amplitude of the density correlation function (multiplied by
$2^{1/2-K}$).   
Finally, we may readily generalize Eq.
(\ref{frie-inf}) to the case of a finite chain of length $L$, by a
standard conformal transformation:
\begin{equation}
<n_j>\to {2\sqrt{c}|A|\cos (2\pi nj + \beta )\over [(2L/\pi )\sin
(\pi j/L)]^{K_{+\rho}}}.\label{fried-fin}\end{equation} The decay
of Friedel oscillations allows a way of determining the critical
exponent $K_{+\rho}$ which is alternative to measuring directly
the long distance behavior of the density correlations.  Indeed
this latter measurement becomes quite difficult with open b.c.'s
due to the necessity of eliminating boundary effects.  

We  
emphasize that the wave-vector of these Friedel oscillations
$2(k_{Fe}+k_{Fo})=2\pi n$, or equivalently $2\pi x$ where 
$x\equiv 1-n$ is the hole density, is a characteristic 
feature of the particular C1S0 phase that we are assuming. 
More correctly, this is the {\it minimum} Friedel oscillation 
wave-vector since higher harmonics are also expected to occur.  
This minimum oscillation wave-vector would be different in 
a different phase.  This wave-vector corresponds to 
2 holes per wave-length and is the same wave-vector that 
would occur for a 1-component spinless hard core bose gas, 
which is an approximate description of the C1S0 phase in 
which hole pairs are assumed to form tightly bound rung 
singlets.\cite{Siller}

\begin{figure}
\includegraphics[width=8cm]{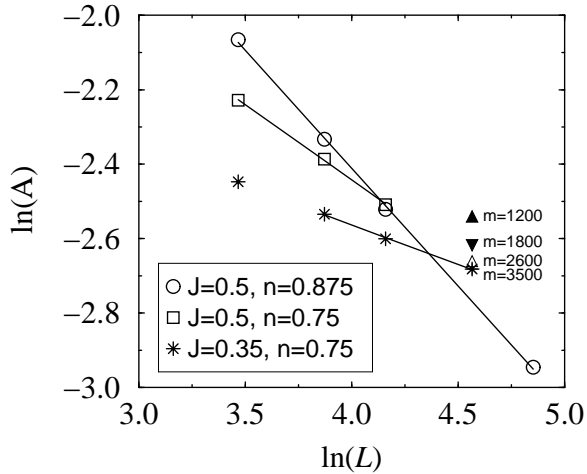}
\caption{The amplitude, $A$, of the Friedel oscillations in the 
center of the system, as a function of the length, $L$. The resulting 
slopes are $-K_{+\rho}$ and are given in Table \ref{table:v,K}. The 
triangles show results for $L=64$ as a function of the number of 
block states kept, $m$, for $J=.35$, $n=.75$.  One can see the 
very slow convergence of the amplitude with the number of states kept.
}
\label{figampall}
\end{figure}

An immediate consequence of Eq. (\ref{fried-fin}) is that 
the amplitude of the density oscillations near the center of a finite 
chain scales as $L^{-K_{+\rho}}$. We show a log-log plot of this amplitude
versus  length in Fig. \ref{figampall}.  Fitting to a 
straight line allows a determination of $K_{+\rho}$.  The 
corresponding values of $K_{+\rho}$ determined in this way 
are shown in Table \ref{table:v,K}.  We see that they 
are roughly comparable to the values obtained from the finite 
size spectrum.  

\begin{figure}[b]
\includegraphics[width=8cm]{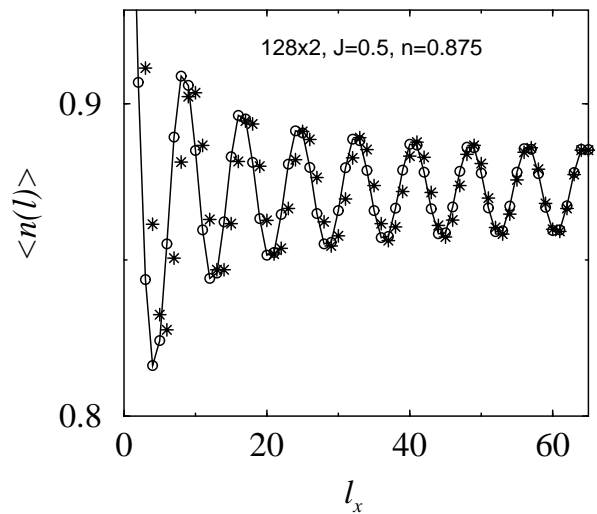}
\caption{Density at site $l$ from DMRG (* symbols) compared 
to Eq. (\ref{fried-fin}) (circles and lines)for $(n,J)=(.5,.875)$
 using 
$K_{+\rho}=.63$.
}
\label{fig:n_j.5}
\end{figure}

\begin{figure}
\includegraphics[width=8cm]{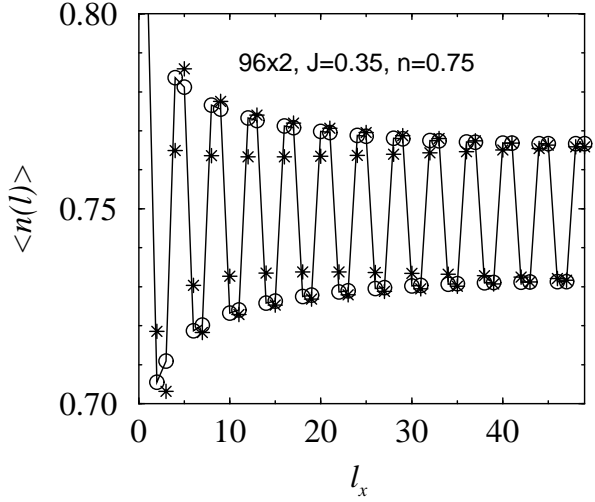}
\caption{Density at site $l$ from DMRG (* symbols) for $(n,J)=(.35,.75)$
compared 
to Eq. (\ref{fried-fin}) (circles and lines) using 
$K_{+\rho}=0.33$.
}
\label{fig:n_j.35}
\end{figure}

In Figs.   \ref{fig:n_j.5} and \ref{fig:n_j.35} we show 
the Friedel oscillations in the density at site $l$, for 
two different values of $(n,J)$, fitted to 
Eq. (\ref{fried-fin}) with  $K_{+\rho}$ 
taken as a free parameter. 
The agreement is fair although the presence of corrections 
due to irrelevant operator effects is evident.
 In both cases the value of 
$K_{+\rho}$ so determined is in rough agreement with 
the values in the third column of Table I, determined from 
the finite size spectrum.  

We emphasize that these formulas are true very generally for
Luttinger liquids with a single gapless charge boson.  In
particular, they apply to the spinless single chain model.  In the
non-interacting case we may readily find the exact formula for the
Friedel oscillations.  For $N$ electrons on $L$ sites:
\begin{eqnarray}
&&<n_j>={2\over L+1}\sum_{m=1}^N\sin^2{\pi mj\over L+1}
\nonumber \\
&&={N+1/2\over
L+1}-{\sin [ 2\pi j(N+1/2)/(L+1)] \over 2(L+1)\sin [(\pi
j/(L+1)]}.\end{eqnarray} 
For large $N$ and $L$ this can be approximated:
\begin{equation}
<n_j>\approx n-{\sin 2\pi nj \over 2L\sin (\pi
j/L)}.\label{free}\end{equation} This has the expected form of Eq.
(\ref{fried-fin}) with $K=1$ and $\sqrt{c}|A|=1/2\pi$. On the
other hand, the $2k_F$ part of the density correlation function at
long distances is:
\begin{equation}
<n_jn_0>\to {\cos 2\pi nj\over 2\pi^2|j|^2}\end{equation} which has
the form of Eq. (\ref{dens-corr}) with the same value of
$\sqrt{c}|A|=1/2\pi$.

We see that, not surprisingly, when the charge density
correlations drop off slowly, so do the Friedel oscillations.  In
particular, this makes it difficult to determine numerically
whether or not a CDW occurs at $n=3/4$, for example, by measuring
density oscillations.  If $J/t$ is such that the system almost has
a CDW then $K_{+\rho}$ will be only slightly greater than 1/8.  The
extremely slow decay of the Friedel oscillations will be difficult
to distinguish from a true CDW, where the oscillation amplitude
goes to a non-zero constant far from the boundaries.

We now consider the single chain Hubbard and $tJ$ models.  The 
Hubbard model is known to be in a C1S1 phase for all densities 
(except $n=1$) and all $U>0$.  Thus the low energy 
effective Hamiltonian contains both spin and charge bosons.  (We use 
the same notation as for the 2-leg ladder except that the $\pm $ 
labels are no longer needed since there is only one type of charge 
boson and one type of spin boson.) The charge boson Hamiltonian 
is written exactly as in Eq. (\ref{Ham}), in terms of the parameter 
$K_{\rho}$.  The spin boson Hamiltonian also has exactly this form 
but with $K_{\sigma}=1$, 
as follows from $SU(2)$ spin rotation invariance.  The 
pairing, $2k_F$ and $4k_F$ density operators now take the form:
\begin{eqnarray}
\psi_{L\uparrow}\psi_{R\downarrow}& \propto& e^{i\sqrt{2\pi}(\phi_{\rho}
-\theta_{\sigma})}\nonumber \\
e^{-2ik_Fx}\psi^\dagger_{R\uparrow}\psi_{L\uparrow} &\propto & 
e^{-2ik_Fx}e^{-i\sqrt{2\pi}(\theta_{\rho}+\theta_{\sigma})}\nonumber \\
e^{-4ik_Fx}\psi^\dagger_{R\uparrow}\psi_{L\uparrow}
\psi^\dagger_{R\downarrow}\psi_{L\downarrow}&\propto &
e^{-4ik_Fx}e^{-2i\sqrt{2\pi}\theta_{\rho}}.
\end{eqnarray}
Note that, in this case, we have formed a $4k_F$ operator 
in which the spin boson doesn't appear.  It then follows 
that the pair correlations, $2k_F$ density correlations 
and $4k_F$ density correlations decay with power law 
exponents:
\begin{eqnarray}
\eta_{\hbox{pair}}&=&1+1/K_{\rho}\nonumber \\
\eta_{2k_F}&=& 1+ K_{\rho} \nonumber \\
\eta_{4k_F} &=& 4K_{\rho}.\end{eqnarray}
Open b.c.'s imply:
\begin{eqnarray}
\theta_{\rho}(0)&=&\hbox{constant}\nonumber \\
\theta_{\sigma}(0)&=& \hbox{constant}.\end{eqnarray}
It thus follows that the density will exhibit both $2k_F$ and
 $4k_F$ Friedel oscillations with different exponents:
\begin{equation}
<n_j> \to n +{A \cos (\pi nj+\alpha )\over j^{(1+K_{\rho})/2}}
+{B\cos (2\pi nj+\beta )\over j^{2K_{\rho}}},\end{equation}
Where we have expressed the oscillation wave-vectors in terms of the 
electron density using the  free electron result for electrons 
with spin on a single chain:
\begin{equation}
2k_F=\pi n.\end{equation}
The amplitudes, $A$ and $B$ are proportional to the square 
roots of the corresponding terms in the density correlation 
function and depend on density and $U$ (as does $K_{\rho})$. 

\begin{figure}
\includegraphics[width=8cm]{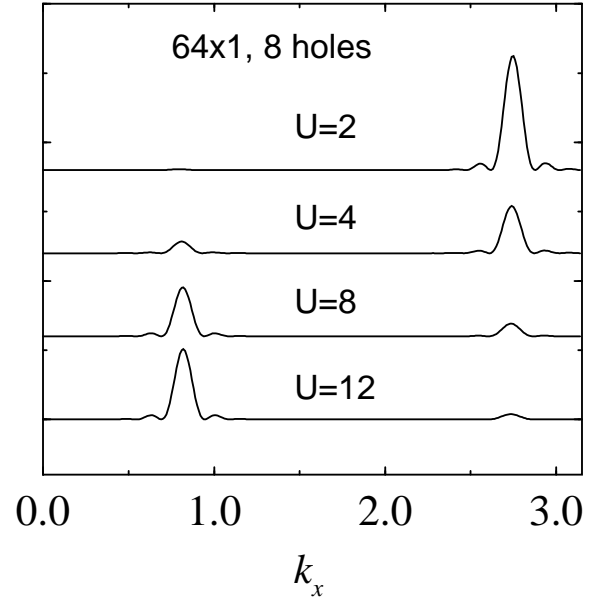}
\caption{Smoothed Fourier transform of the density, $n_j$ in the 
single chain Hubbard model at 7/8 filling for various values of 
the repulsion strength, $U$.}
\label{figallU}
\end{figure} 

It is interesting to consider the limit $U>>t$ where the no double 
occupancy constraint is present.  If we consider 
some initial configuration of holons and spins, then the order 
of the spins along the chain can never change under nearest neighbor
  hopping 
processes consistent with no double occupancy.  It follows that 
the charge dynamics must be identical to those of a system 
of {\it non-interacting}
 spinless fermions with  the same density.  That is to say, 
the charge dynamics is ``spin blind''.  It follows that 
the density oscillations should not have the $\pi n$ term in this 
limit but only the $2\pi n$ term as for spinless fermions 
[Eq. (\ref{free})]; 
i.e. that $A\to 0$ at $U\to \infty$.
This argument also determines $K_{\rho}=1/2$ and $B=1/2\pi$ in 
this limit.  We remark that this argument is special to the 
single chain model with only nearest neighbor hopping.  
On a ladder, or even on a chain with next nearest neigbor hopping, 
 spin rearrangement is possible 
without double occupancy by moving the electrons around  
or past each other.   

In Fig. \ref{figallU} we plot the Fourier transform of the 
density  
for the Hubbard model, at 7/8 filling, for various values of $U$. 
The same smoothing procedure was used that we mentioned in Sec. III.   
In this case $2k_F=\pi n = 7\pi /8$ and $4k_F=7\pi /4$ or equivalently 
$\pi /4$.  
Both $2k_F$ and $4k_F$ components are clearly visible and it is 
evident that the $2k_F$ component vanishes at large $U$.

\begin{figure}[b]
\includegraphics[width=8cm]{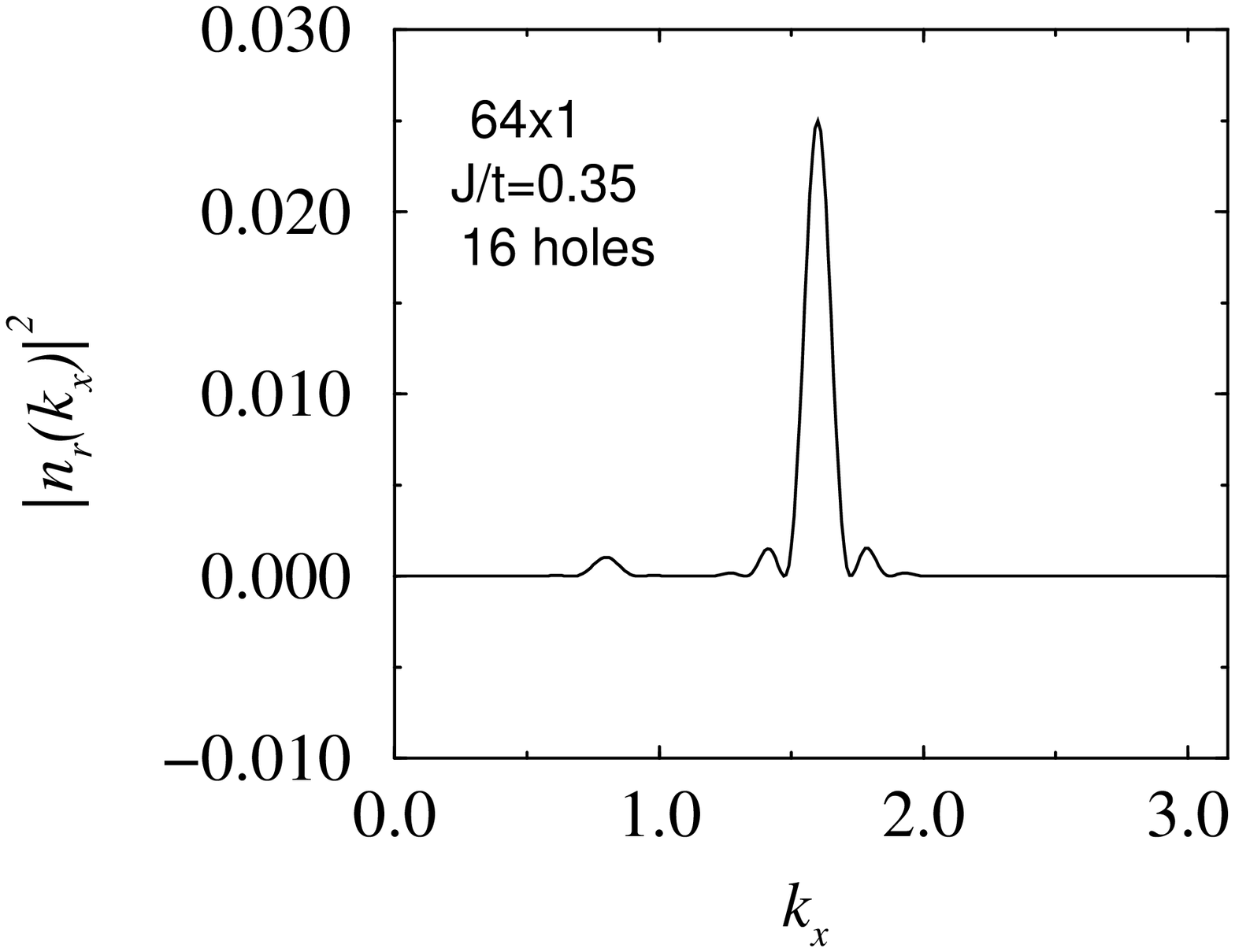}
\includegraphics[width=8cm]{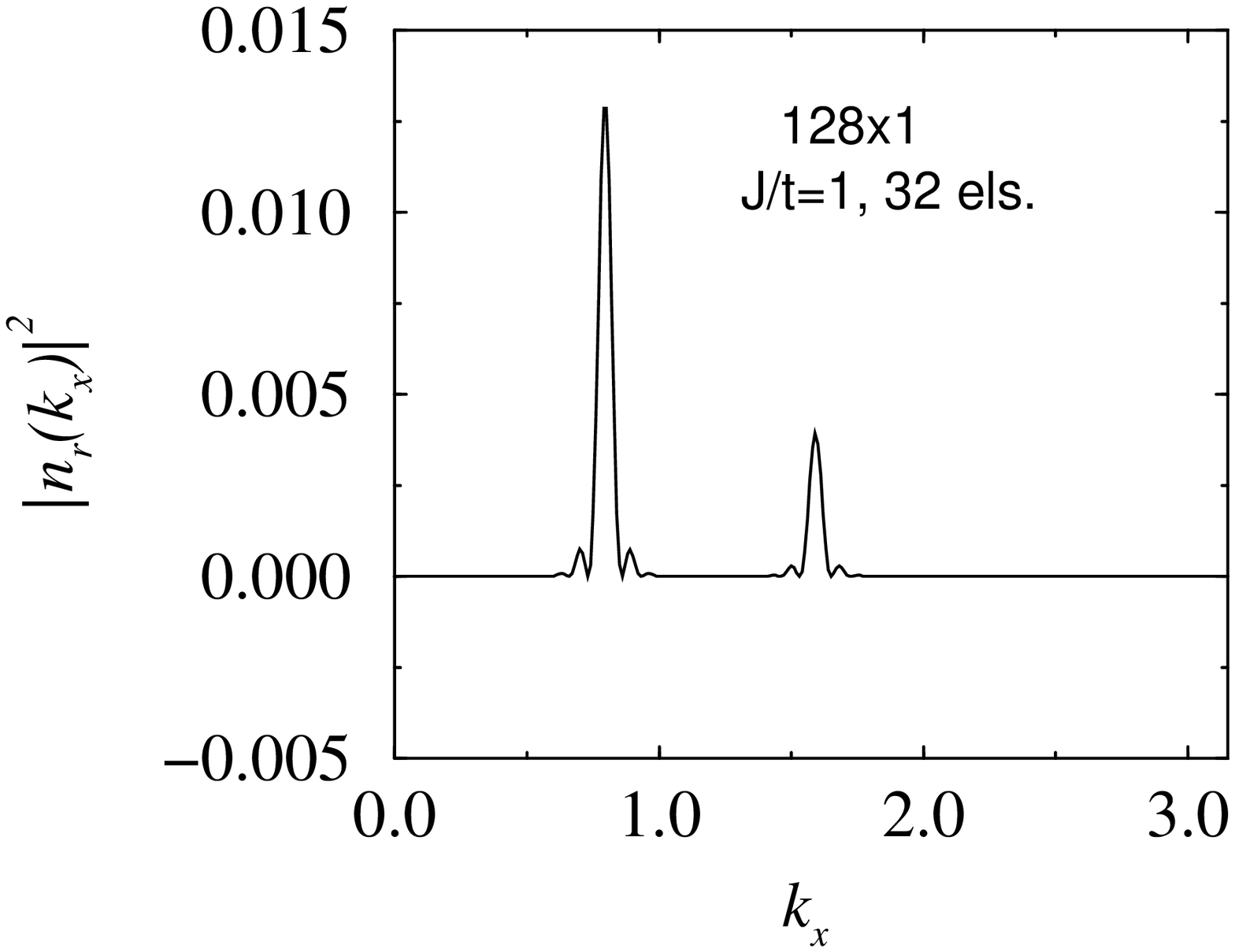}
\caption{Smoothed Fourier transform of the density, $n_j$ in the 
single chain $tJ$ model at 1/4 filling for a) $J/t=.35$ and b) J=1.0, 
showing Friedel oscillations at $2k_F=\pi /4$ and $4k_F=\pi /2$.}
\label{fig:1legtJ}
\end{figure} 

What does this imply about the (single chain) $t-J$ model?
  The $J\to 0$ limit 
of the $t-J$ model is identical to the $U\to \infty$ limit 
of the Hubbard model.  Therefore, in that limit, the density 
oscillations have vanishing $2k_F$ component.  It is not 
completely obvious  what happens at non-zero $J$.  Insofar 
as this is the same as large but finite $U$, we would expect 
to have a small $2k_F$ component to the density oscillations.  
Note however, that the no double occupancy constraint is enforced 
exactly in the $t-J$ model, by construction, whereas it is 
not in the finite $U$ Hubbard model.  Thus we must understand 
whether it is the no double occupancy which is responsible for 
the vanishing $2k_F$ oscillations or whether the vanishing 
of all spin rearrangement processes is neccessary.  The argument in the 
previous paragraph seems to rely on the vanishing of spin rearrangement 
processes so we
suspect that the Friedel oscillations should exhibit a 
$2k_F$ component for any finite $J$.  Of course, for small 
$J/t$ this is expected to be small.  In Fig.  
\ref{fig:1legtJ}  we show the Fourier transformed 
density for the single leg $t-J$ model at density $n=1/4$ 
for two values of $J/t$.  Note that now $2k_F=\pi /4$ and 
$4k_F=\pi /2$.  For smaller $J/t$ the $4k_F$ oscillations 
clearly dominate although a small $2k_F$ part is observed.  
At larger $J/t$, the oscillations at $2k_F$  are larger than 
those at $4k_F$.  

Thus we expect $tJ$ models to exhibit generic Friedel oscillations in 
all cases except for the special case of a single leg in the 
limit $J/t\to 0$ where the absence of spin rearrangement processes 
eliminates the lowest wave-vector component of the  oscillations.  

The Friedel oscillation wave-vector can be a useful diagnostic of 
which phase a particular multi-leg $tJ$ model is in.  For instance, 
the minimum oscillation wave-vector, $2k_{Fe}+2k_{Fo} =2\pi n$,
is a characteristic of the particular C1S0 phase of the 2 leg ladder 
reviewed in Sec. II (in which the $\theta_{\pm \sigma}$ 
and $\phi_{-\rho}$ fields are pinned).  Other phases exhibit other 
oscillation wave-vectors.  We expect such an analysis of DMRG 
results to be useful 
in determining the phase diagram of multi-leg ladders.

\acknowledgements DJS would like to thank T. Giamarchi 
for helpful discussions.  
DJS and IA acknowlege the hospitality of the ITP, 
University of California, Santa Barbara where this work was 
initiated.  SRW acknowledges support from the NSF under 
grant No. DMR98-70930; IA acknowleges support from 
the NSF Grant No. PHY99-07949 (ITP) 
and NSERC of Canada; DJS acknowledges support from the NSF under grant 
No. PHY99-07949 (ITP) and grant No. DRM98-17242 (DJS).


\begin{thebibliography}{999}
\bibitem{Fabrizio} M. Fabrizio, Phys. Rev. {\bf B37}, 325 (1993).
\bibitem{Schulz} H.J. Schulz, Phys. Rev. {\bf B53}, 2959 (1996).
\bibitem{Balents} L. Balents and M.P.A. Fisher, Phys. Rev. 
{\bf B53}, 12133 (1996).
\bibitem{Orignac} E. Orignac and T. Giamarchi, Phys. Rev. 
{\bf B56}, 7167 (1997).
\bibitem{Noack} R.M. Noack, S.R. White and D.J. Scalapino,
 Phys. Rev. Lett. {\bf 73}, 882 (1994).
\bibitem{Siller} T. Siller, M. Troyer, T.M. Rice and S.R. White, 
Phys. Rev. {\bf B63}, 195106 (2001).
\bibitem{Hayward} C.A. Hayward and D. Poilblanc, Phys. Rev. 
{\bf B53}, 11721 (1996).
\bibitem{Eggert} S. Eggert and I. Affleck, Phys. Rev. Lett. {\bf 75}, 
934 (1995).
\bibitem{Egger} R. Egger and H. Grabert, Phys. Rev. Lett. {\bf 75}, 
3505 (1995). 
\bibitem{White2} S.R. White and D.J. Scalapino, Phys. Rev. {\bf B61}, 
6320 (2000).
\bibitem{Schulz2} H.J. Schulz, Phys. Rev. {\bf B59}, 2471 (1999).
\bibitem{Schulz3} H.J. Schulz, Phys. Rev. {\bf B22}, 5274 (1980).
\bibitem{Giamarchi}, T. Giamarchi, Physica {\bf B230}, 975 (1997).  
\bibitem{Eggert2} S. Eggert and I. Affleck, Phys. Rev. {\bf B46}, 
10866 (1992).
\bibitem{Yamanaka} M Yamanaka, M. Oshikawa and I. Affleck, Phys. 
Rev. Lett. {\bf 79}, 1110 (1997).
\bibitem{smooth} M. Vekic and S.R. White, Phys. Rev. Lett. {\bf 71},
4283 (1993); see Eq. (4) and (10). 
\bibitem{OAF} D.J. Scalapino, S.R. White and I. Affleck, Phys. Rev. 
{\bf B64}, 100506 (2001).  J.B. Marston and A. Sudbo, cond-mat/0103120.
\bibitem{separation} S. Rommer, S.R. White and D.J. Scalapino, 
Phys. Rev. {\bf B61}, 14701 (1997).
\end{thebibliography}
\end{document}